\documentclass[12pt]{article}

\usepackage[english,russian]{babel}
\usepackage{amsfonts}
\usepackage{amsmath}
\usepackage{amssymb}
\usepackage{latexsym}

\newcommand{\Aut}{{\rm Aut}}

\newcommand{\rk}{{\rm rk}\,}
\newcommand{\id}{{\rm id}\,}

\newcommand{\ov}{\overline}
\newcommand{\wt}{\widetilde}
\newcommand{\ot}{\otimes}
\newcommand{\la}{\langle}
\newcommand{\ra}{\rangle}
\newcommand{\lra}{\longrightarrow}
\newcommand{\sdir}{\leftthreetimes}
\newcommand{\col}{\colon}
\newcommand{\sse}{\subseteq}
\newcommand{\fe}{\varphi}
\newcommand{\mC}{{\mathbb C}}
\newcommand{\cA}{{\mathcal A}}
\newcommand{\cB}{{\mathcal B}}
\newcommand{\cO}{{\mathcal O}}
\newcommand{\cP}{{\mathcal P}}
\newcommand{\cT}{{\mathcal T}}

\begin{document}

\begin{center} \large On automorphism group of a possible short algorithm \\ 
for multiplication of  $3\times3$ matrices \\
\medskip
\normalsize Vladimir~P.~Burichenko  \\
\medskip
\small Institute of Mathematics of the National Academy of Sciences of Belarus \\
 e-mail: vpburich@gmail.com 
\end{center}
\bigskip

\begin{flushright}
\em To the memory of Irina Dmitrievna Suprunenko
\end{flushright}
\medskip

\begin{abstract}
Studying algorithms admitting nontrivial symmetries is a prospective way of constructing 
new short algorithms of matrix multiplication. The main result of the article is that if there 
exists an algorithm of multiplicative length $l\leq22$ for multuplication of $3\times3$ matrices 
then its automorphism group is isomorphic to a subgroup of $S_l\times S_3$. 
\end{abstract}

\paragraph{1. Introduction. Basic definitions and the result.}  
Fast matrix multiplication is one of the main issues in the complexity theory. This question may 
be easily stated in the language of tensor decompositions. Recall necessary definitions. 

Let $\wt V=V_1\ot\ldots\ot V_l$ be the tensor product of several finite dimensional vector spaces. 
A tensor of the form $v_1\ot\ldots\ot v_l$ is called {\em elementary} 
or {\em decomposable.} An (additive) {\em decomposition} of length $s$ for $w$ is any 
unordered set $\{w_i\mid i=1,\ldots,s\}$ of elementary tensors such that $\sum_{i=1}^s w_i=w$. 

The {\em rank} $\rk(w)$ is the minimal possible decomposition length for $w$. 

Let    
$$M_{mn}=M_{m,n}(\mC)=\la e_{ij}\mid 1\leq i\leq m, \ 1\leq j\leq n\ra_\mC$$
be the space of all complex $m\times n$ matrices, where $e_{ij}$ are the usual matrix unities. 
Consider the following tensor in the space $M_{mn}\ot M_{np}\ot M_{pm}$: 
$$ \cT=\sum_{1\leq i\leq m,\, 1\leq j\leq n,\, 1\leq k\leq p} e_{ij}\ot e_{jk}\ot e_{ki}\,.$$
It is often denoted also by $\la m,n,p\ra$, cf. \cite[Section 14.2]{BCS}. 

The question of main interest in the algebraic complexity theory is the rank 
of $\cT$ (or at least the estimates for this rank), because of the following fact 
(see \cite[Section 15.3]{BCS}). 
\smallskip

\noindent{\bf Proposition 1.} {\em Assume that for some $m,n,p\geq1$ the tensor $\la m,n,p\ra$ 
has a decomposition of length $l$. Then there exists an algorithm for multiplication of 
$N\times N$ matrices over an arbitrary field of characteristic 0 of complexity $O(N^\tau)$ arithmetical 
operations, where $\tau=(3\ln l)/\ln mnp$. }

\smallskip

The following estimates for $\rk(\la m,n,p\ra)$ are known: $\rk(\la 2,2,2\ra)=7$ (here the upper 
estimate follows from the Strassen algorithm, and the lower one is well-known also); 
$\rk(\la 2,2,3\ra)=11$, \cite{Alexeev}; $\rk(\la 2,3,3\ra)\in\{14,15\}$ \cite{Hop-Kerr}, \cite{Blaser1}; 
$19\leq\rk(\la 3,3,3\ra)\leq23$ \cite{Laderman}, \cite{Blaser2}. 
Some estimates for other $\rk(\la m,n,p\ra)$ are known also, but we shall not need them. 
It should be mentioned that $\rk(\la m,n,p\ra)$ is invariant under permutations of $m,n,p$, 
and, finally, that always $\rk(\la m,n,1\ra)=mn$.  

The most appealing in this field is the following question: what is the rank of $\la3,3,3\ra$? 
It was found, by using numerical methods, a lot of new decompositions of lenght $23$, 
but no one of length $\leq22$. See \cite{JM}, \cite{Bard_ea}, \cite{3korejca}, \cite{Smirnov}, 
\cite{HKS1}, \cite{HKS2}. 

One of prospective ways of search for short decompositions of $\la m,n,p\ra$ 
is to study decompositions admitting nontrivial symmetry groups. This approach was proposed by 
the author in preprints \cite{Bur2014}, \cite{Bur2015} and independently by Landsberg and co-authors 
in \cite{CILO2016}, \cite{CILO}, \cite{Landsberg2018}. 

Recall the definitions related to decompositions automorphisms.  
Let $\wt V=V_1\ot\ldots\ot V_l$ be as above. To avoid long formulae, we consider only the case 
$l=3$. Let $S(\wt V)$ be the group of all nondegenerate linear transformation of $\wt V$ that 
preserve the tensor decomposition of this space, but possibly permuting the factors. 
For example, the transformations of the form   
$$ v_1\ot v_2\ot v_3\mapsto \alpha(v_2)\ot\beta(v_1)\ot\gamma(v_3)\,,$$
where $\alpha\col V_2\lra V_1$ and $\beta\col V_1\lra V_2$ are isomorphisms, and 
$\gamma$ is a nondegenerate transformation of $V_3$ (note that in this case it is necessary to 
suppose that $\dim V_1=\dim V_2$). Obviously, $S(\wt V)$ preserves the set of 
decomposable tensors. (Sometimes the elements of $S(\wt V)$ are called {\em Segre automorphisms}, 
because when acting on the projectivization $(\wt V\setminus\{0\})/\mC^\ast$ they preserve the 
Segre variety (the image in the projective space of the set of all nonzero decomposable tensors)). 
The subgroup of elements of $S(\wt V)$, corresponding to the trivial permutation of the factors, 
that is of the form $A\ot B\ot C$, where $A\in GL(V_1)$, $B\in GL(V_2)$, and $C\in GL(V_3)$, 
will be denoted by $S^0(\wt V)$. 

For a tensor $w\in\wt V$ call   
$$ \Gamma(w)=\{g\in S(\wt V)\mid gw=w\}$$
{\em the isotropy group} of $w$, and the intersection $\Gamma^0(w)=\Gamma(w)\cap S^0(\wt V)$ 
--- the {\em small} isotropy group. 

Let $\cP=\{w_1,\ldots,w_l\}$ be a decomposition for $w$. Consider $\cP$ as a multiset, 
that is an unordered set some of whose elements may be equal.  
The {\em automorphism group} for $\cP$ is the subgroup of all elements of $S(\wt V)$ preserving $\cP$: 
$$ \Aut(\cP):=\{g\in S(\wt V)\mid g\cP=\cP \}.$$ 
It is clear that $\Aut(\cP)\leq \Gamma(w)$. 

The isotropy group of $\la m,n,p\ra$ can be easily described (but the proof is not trivial !). 
Such a description was obtained in \cite{deGroote1}, \cite{Br-Dob}, and more accurately in  
\cite{Bur2016}. Specifically, $\Gamma^0(\la m,n,p\ra)$ consists of all transformations 
of $M_{mn}\ot M_{np}\ot M_{pm}$ of the form 
$$ T(a,b,c)\col x\ot y\ot z\mapsto axb^{-1}\ot byc^{-1}\ot cza^{-1}\,,$$
where $a\in GL(m,\mC)$, $b\in GL(n,\mC)$, and $c\in GL(p,\mC)$ (and observe that $T(\lambda a, 
\mu b,\nu c)=T(a,b,c)$, where $\lambda,\mu,\nu\in\mC^\ast$). If $m,n,p$ are pairwise distinct, 
then $\Gamma(\la m,n,p\ra)$ coincides with $\Gamma^0$, whereas if some of $m,n,p$ are equal, 
then $\Gamma$ is the semidirect product $\Gamma(\la m,n,p\ra)=\Gamma^0(\la m,n,p\ra)\sdir Q$, where 
$Q\cong Z_2$, if two of $m,n$,and $p$ are equal, and $Q\cong S_3$ if $m=n=p$. In the latter case 
we can take $Q=\la\sigma,\rho\ra$, where  
$$ \sigma\col x\ot y\ot z\mapsto z\ot x\ot y\,,\qquad \rho\col x\ot y\ot z\mapsto 
y^t\ot x^t\ot z^t $$
(here $x^t$ is the matrix transpose to $x$). 

The following approach is rather natural: take a subgroup $G\leq\Gamma(\cT)$, where $\cT=\la3,3,3\ra$,
and study the $G$-invariant decompositions of $\cT$, trying to find a decomposition of 
length $\leq22$. This was done in \cite{Bur2021}, \cite{Bur2022} for a certain 
(rather large) subgroup, isomorphic to $S_4\times S_3$, and in \cite{Chokaev-eng} and 
\cite{Landsberg2018} for a certain subgroup $\cong Z_3$. In the former case it was shown that 
a $G$-invariant decomposition of length $\leq23$ does not exist. In the latter case two 
$G$-invariant decompositions of length $25$ were found (in \cite{Chokaev-eng}), and a new 
decomposition of length $23$ in \cite{Landsberg2018}, whose full automorphism group happens to  
be isomorphic to $Z_4\times Z_3$. 

Obviously, the group $\Gamma(\cT)$ contains infinitely many subgroups (even up to cojugacy). 
A natural question arises, can we restrict {\em in advance} the set of subgroups $G$ that can appear, 
in principle, as an automorphism group of a possible decomposition of length $\leq22$ (if such 
a decomposition do exists ?). To obtain such a restricting condition is the aim of the present work. 
The following theorem is proved. 
\smallskip

\noindent{\bf Theorem 2.} {\em  Let $\cP=\{w_1,\ldots,w_l\}$ be a decomposition 
for $\la3,3,3\ra$ of length $l\leq22$. Then $\Aut(\cP)$ is isomorphic to a subgroup 
of $S_l\times S_3$. More precisely, the map $g\mapsto(\alpha(g),\beta(g))$ is injective, 
where $\alpha(g)$ is the permutation of elements of $\cP$, induced by $g\in\Aut(\cP)$, and 
$\beta(g)$ is the permutation of the factors of $M_{33}^{\ot3}$, corresponding to $g$.  }
\medskip

\paragraph{2. A direct sum of tensors.} If $V_1$, $V_2$, and $V_3$ are three spaces and 
$V'_i\sse V_i$ are their subspaces, then the tensor product $\wt V'=V'_1\ot V'_2\ot V'_3$ 
is naturally considered as a subspace of $\wt V=V_1\ot V_2\ot V_3$. It is easy to see that 
for any tensor $w\in\wt V'$ its rank with respect to the decomposition $V'_1\ot V'_2\ot V'_3$ 
is the same as the rank with respect to $V_1\ot V_2\ot V_3$. 

Now assume that the spaces $V_i$ are decomposed into direct sums: $V_i=V'_i\oplus V''_i$, 
and $w'$ and $w''$ are the elements of $V'_1\ot V'_2\ot V'_3$ and $V''_1\ot V''_2\ot V''_3$, 
respectively. Then we can consider the tensor $w=w'+w''$. In such a situation it is said that 
$w$ is a {\em direct sum} of $w'$ and $w''$, $w=w'\oplus w''$. (Strictly speaking, we should 
say this is a {\em inner} direct sum, similarly to concept of inner direct sum of space or inner direct 
product of groups. We can also define an outer direct sum of tensors, in an evident way.)

Obviously, $\rk(w)\leq\rk(w')+\rk(w'')$. There was a long-standing conjecture (the {\em 
Strassen direct sum conjecture}) that the latter inequality is actually always an equality. 
It was however failed, see \cite{Shitov}. Nevertheless, the direct sum conjecture is true in 
a certain particular case. 
\smallskip

\noindent{\bf Proposition 3.} {\em Suppose that at least one of the six spaces $V'_i$, $V''_i$ 
is of dimension $\leq2$. Then $\rk(w'\oplus w'')=\rk(w')+\rk(w'')$. }
\smallskip

{\em Proof.} See \cite{JaJa}. Another proof is contained in \cite{Buczynski}. 
\hfill $\square$ \smallskip

We also need another two general concepts regarding tensors. 

1) We say that $w\in V_1\ot V_2\ot V_3$ and $w'\in V'_1\ot V'_2\ot V'_3$ are 
{\em similar}, and denote this by $w\sim w'$, if there exist isomorphisms $\fe_i\col V_i\lra V'_i$ 
such that $(\fe_1\ot\fe_2\ot\fe_3)w=w'$. Similarly, $w$ and $w'$ are {\em similar up to a permutation
of tensor factors}, if there exist a permutation $\pi\in S_3$ and isomorphisms $\fe_i\col V_i
\lra V'_{\pi i}$ such that $\wt\fe w=w'$, where $\wt\fe=\pi\circ(\fe_1\ot\fe_2\ot\fe_3)$. 

Clearly, if two tensors are similar, or similar up to a permutation of factors, 
then they are of the same rank. 

2) Notice that for any two subspaces $V'_1,V''_1\sse V_1$ we have     
$$ (V'_1\ot V_2\ot V_3)\cap(V''_1\ot V_2\ot V_3)=(V'_1\cap V''_1)\ot V_2\ot V_3\,.$$
It follows that if $\{w_\alpha\in V_1\ot V_2\ot V_3\mid \alpha\in\mathrm A\}$ is any family 
of tensors, the there exists the least subspace $U\sse V_1$ such that 
$U\ot V_2\ot V_3$ contains all $w_\alpha$. This $U$ will be called {\em the (tensor) projection} 
of the family $\{w_\alpha\}$ to $V_1$. The projections to other factors are defined similarly. 

It is easy to see that the projection of $\{w_\alpha\}$ to $V_1$ is nothing else but the span 
of convolutions of the tensors $w_\alpha$ with all tensors of the form $l_2\ot l_3$, 
where $l_2\in V_2^\ast$, $l_3\in V_3^\ast$. So it is easy to find the projection of 
$\cT$ to each of the three factors $M$: it is the whole $M$.

\paragraph{3. Some identifications.}  
Let $V=\mC^3$ be the space of all column vectors of height $3$. Its dual space  
$V^\ast$ may be identified with the space of rows of length $3$, so that $\la l,v\ra=l(v)=lv$, 
where $l$ is a row, $v$ a column (note that $lv$ is a $1\times1$ matrix, that is, a number). 

The group $GL(V)=GL(3,\mC)$ acts (on the left) on $V$ as usually, i.e., $g(v)=gv$. This group 
acts (on the left) also on $V^\ast$ by $g(l)=lg^{-1}$. This actions are compatible in the sense 
that always $\la g(l),g(v)\ra=\la lg^{-1},gv\ra=(lg^{-1})(gv)=lv=\la l,v\ra$. 

Next, the space $V\ot V^\ast$ can be identified with $M=M_3(\mC)$ by the rule  
$v\ot l\mapsto vl$ (note that the product of column by a row is a $3\times3$ matrix). 
The matrix corresponding to the tensor $e_i\otimes e^j$ under this identification is 
the matrix unit $e_{ij}$ (here $e_i$ and $e^i$ are the column and the row, respectively, 
that have $1$ in $i$-th position, and $0$ in other places). 

The space $V\ot V^\ast$ is acted on by the group $GL(V)\times GL(V)$ by the rule  
$$ (g_1,g_2) (v\ot l)=g_1(v)\ot g_2(l)=g_1v\ot lg_2^{-1}\,.$$
When $V\ot V^\ast$ identifies with $M$, the corresponding action on $M$ is 
$$ (g_1,g_2)(x)=g_1xg_2^{-1}\,. $$

Let $V=V_1\oplus\ldots\oplus V_k$ be a decomposition into a direct sum. Let  
$L_i$ be the subspace in the row space $V^\ast$, consisting of all $l$'s such that  
$\la l,v\ra=0$ for all $v\in V_j$, $j\ne i$. Then it is easy to see that $V^\ast=L_1\oplus
\ldots\oplus L_k$ and the pairing of $V_i$ and $L_i$ is nondegenerate, so that $L_i$ 
is identified with $V_i^\ast$, and so $V^\ast$ identifies with $V_1^\ast\oplus\ldots\oplus V_k^\ast$. 

We apply the identifications described to prove the following. 
\smallskip

\noindent{\bf Proposition 4.} {\em Let $V=V_1\oplus\ldots\oplus V_m=U_1\oplus\ldots\oplus U_s$ 
be two decompositions for $V$, let $k\leq\min(m,s)$ and $K=\bigoplus_{i=1}^k V_i\ot U_i^\ast$, 
and let $N$ be the sum of all the remaining summands $V_i\ot U_j^\ast$, so that 
$M=V\ot V^\ast=K\oplus N$. Let $\zeta$ and  $\xi$ be the components of $\cT$ in 
$K\ot M\ot M$ and $N\ot M\ot M$, respectively. Then 
$$ \zeta\sim \bigoplus_{i=1}^k \la p_i,q_i,3\ra\,,$$
where $p_i=\dim V_i$ and $q_i=\dim U_i$. Moreover, $\rk\zeta=\sum_{i=1}^k\rk(\la p_i,q_i,3\ra)$. }
\smallskip

{\em Proof}. First we reduce the statement to the particular case where $V_i$ and $U_j$ 
are coordinate subspaces, that is  $V_i=\la e_\alpha\mid\alpha\in I_i\ra_\mC$, 
$U_i=\la e_\alpha\mid\alpha\in J_i\ra_\mC$, where $I_1,\ldots,I_m$ are disjoint subsets 
that form a partition of $\{1,2,3\}$, as well as $J_i$. 

Obviously, there exist $a,b\in GL(V)$ such that $aV_i$ and $bU_j$ are coordinate subpaces. 
Then $U_j^\ast b^{-1}=(bU_j)^\ast$  are coordinate subspaces also. 

Consider the transformation $x\mapsto axb^{-1}$ of $M$. The image of $K$ under this transformation 
is     
$$ K_1=aKb^{-1}=\bigoplus_{i=1}^k aV_i\ot U^\ast_i b^{-1}\,, $$
and $N_1=aNb^{-1}$ is the sum of the all remaining $aV_i\ot U^\ast_j b^{-1}$. 

Consider $T=T(a,b,E)$. Then $T$ takes $K\ot M\ot M$ to $K_1\ot M\ot M$, and $N\ot M\ot M$ to  
$N_1\ot M\ot M$. As $T$ preserves $\cT$,  $T$ takes $\zeta$ to $\zeta_1$, 
and $\xi$ to $\xi_1$, where $\zeta_1$ and $\xi_1$ are the components of $\cT$ in $K_1\ot M\ot M$ and 
$N_1\ot M\ot M$, respectively. Clearly,  $\zeta\sim\zeta_1$. As we suppose the proposition 
is true in the case of coordinate subspaces, we see that  $\zeta_1\sim 
\bigoplus_{i=1}^k \la p'_i,q'_i,3\ra$ , where $p'_i=\dim aV_i$, $q'_i=\dim U_i^\ast b^{-1}$. 
But, clearly, $p'_i=p_i$, $q'_i=q_i$. 

Thus, we can assume that $V_i$ and $U_j$ are coordinate subspaces. It is clear that  
$K=\la e_{\alpha\beta}\mid (\alpha,\beta)\in S\ra$, where $S=\sqcup_{i=1}^k I_i\times J_i$, 
and $N=\la e_{\alpha\beta}\mid (\alpha,\beta)\notin S\ra$. Hence it is clear that  
$$ \zeta=\sum_{(\alpha,\beta)\in S,\ 1\leq\gamma\leq3} e_{\alpha\beta}\ot e_{\beta\gamma}
\ot e_{\gamma\alpha}\,.$$
Since the sets $I_i\times J_i$ are disjoint, and the sets $\{1,2,3\}\times I_i$ 
are disjoint also, as well as the sets $J_i\times\{1,2,3\}$, it follows that 
$\zeta=\eta_1\oplus\ldots\oplus \eta_k$, where 
$$ \eta_i=\sum_{\alpha\in I_i,\ \beta\in J_i,\ 1\leq\gamma\leq3} e_{\alpha\beta}
  \ot e_{\beta\gamma}\ot e_{\gamma\alpha}\,.$$
Finally, it is clear that $\eta_i\sim\la|I_i|,|J_i|,3\ra=\la p_i,q_i,3\ra$. This proves the 
first claim. 

Prove the equality for ranks, that is  
$$ \rk \bigoplus_{i=1}^k \la p_i,q_i,3\ra= \sum_{i=1}^k \rk \la p_i,q_i,3\ra\,. \eqno(1)$$
We can assume that the decomposition contains no trivial summands, that is all $p_i,q_i\geq1$. 
If $k=1$, the equality is trivial. If $k\geq3$, then from the inequalities  $p_1+\ldots+p_k\leq3$ 
and $q_1+\ldots+q_k\leq3$ we obtain $k=3$ and $p_i=q_i=1$, $i=1,2,3$. The left-hand side of~(1)
is the direct sum of three copies of the tensor $\la1,1,3\ra$. One of the subspaces 
in the tensor product $M_{11}\ot M_{13}\ot M_{31}$, related to $\la1,1,3\ra$, is one-dimensional, 
so we can apply Proposition 3, whence $\rk 3\cdot \la1,1,3\ra=3\cdot\rk \la1,1,3\ra$ ($=9$). 

It remains to treat the case $k=2$. Again, the inequalities $p_1+p_2\leq3$ and $q_1+q_2\leq3$ imply 
that either at least one of the pairs $(p_1,q_1)$ and $(p_2,q_2)$ is $(1,1)$, or one of these 
pairs is $(1,2)$ and the other $(2,1)$. In the former case at least one of the two summands 
$\la p_i,q_i,3\ra$ is  $\la1,1,3\ra$, and Proposition 3 applies. In the latter case one of the 
summands is $\la1,2,3\ra$, and the other $\la2,1,3\ra$, so Proposition 3 applies again, as 
$M_{1,2}$ is of dimension $2$. 
\hfill $\square$ \medskip

\paragraph{4. The proof of the main theorem.} Now we begin to prove the main theorem. 
Assume on the contrary that the homomorphism $g\mapsto(\alpha(g),\beta(g))$ is not injective. 
This means that there exists $g\in\Aut(\cP)$ which preserves all three of the factors $M$ 
and fixes all tensors $w_i$ of the decomposition $\cP=\{w_i\mid i=1,\ldots,l\}$. Since 
$g$ preserves the factors, it follows that $g=T(a,b,c)$ for some $a,b,c\in GL(3,\mC)$, and 
at least one of $a$, $b$, and $c$ is not a scalar matrix. So we can restate the theorem in the 
following equivalent form: 
\smallskip

\noindent{\bf Proposition 5.} {\em  Let $a,b,c\in GL(3,\mC)$, and at least one of $a$, $b$, 
and $c$ is not a scalar matrix. Suppose that $\{w_i=x_i\ot y_i\ot z_i\mid i=1,\ldots,l\}$ 
is a decomposition for $\cT$ such that all $w_i$ are invariant under $T(a,b,c)$. Then  
$l\geq23$. }
\smallskip

It is this proposition that we are going to prove. 

Notice that we can assume (and will assume below), without loss of generality, that 
$a$ is not scalar. 

We need a lemma. 
\smallskip

\noindent{\bf Lemma 6.} {\em Let $U$ and $V$ be two spaces, $A\in GL(U)$, $B\in GL(V)$. Then 
$A\ot B\in GL(U\ot V)$ is diagonalizable iff both $A$ and $B$ are diagonalizable. }
\smallskip

{\em Proof.} It is obvious that if both $A$ and $B$ are diagonalizable, then $A\ot B$ is 
diagonalizable also. It remains to prove that if one of $A$ and $B$, say $A$, is not 
diagonalizable, then $A\ot B$ is not. There exist $u_1,u_2\in U$ and $\lambda\in
\mC^\ast$ such that $Au_1=\lambda u_1$ and $Au_2=\lambda u_2+u_1$. Also, there exist $v\in V$ 
and $\mu\in\mC^\ast$ such that $Bv=\mu v$. Now we have $(A\ot B)(u_1\ot v)=\lambda\mu(u_1\ot v)$,
$(A\ot B)(u_2\ot v)=Au_2\ot Bv= (\lambda u_2+u_1)\ot (\mu v)=\lambda\mu(u_2\ot v)+\mu(u_1\ot v)$.
Therefore the subspace $\la u_1\ot v,u_2\ot v\ra\sse U\ot V$ is invariant under $A\ot B$ 
and the restriction of  $A\ot B$ to this subspaces is not diagonalizable. 
So $A\ot B$ itself is not diagonalizable. 
\hfill $\square$ \smallskip

\noindent{\bf Statement 7.} {\em Under hypothesis of Proposition 5 both $a$ and $b$ are 
diagonalizable (and $c$ also is). }
\smallskip

{\em Proof.}  Since $T(a,b,c)$ takes $w_i$ to itself, the map $x\mapsto axb^{-1}$ preserves 
all the lines $\la x_i\ra$. It is obvious that the projection of $\cT$ to the first 
factor $M_1$ is contained in the span of all $x_i$. But this projection is $M$, so 
$\la x_i\mid i=1,\ldots,l\ra=M$. So the lines which are invariant under the transformation 
$x\mapsto axb^{-1}$ generate $M$. That is, this transformation of $M$ is diagonalizable. 

When we identify $M$ with $V\ot V^\ast$ this transformation corresponds to $A\ot B$, 
where $Ax=ax$, $Bl=lb^{-1}$. It follows from the lemma that both $A$ and $B$ are diagonalizable. 
So $a$ is diagonalizable, and the transformation $l\mapsto lb^{-1}$ of $V^\ast$ is diagonalizable, 
whence $b$ is diagonalizable also.  
\hfill $\square$ \smallskip

Let $\lambda_1,\ldots,\lambda_s$ be all the distinct eigenvalues of $a$, let $\mu_1,\ldots,
\mu_t$ be the eigenvalues of $b$, and $V=V_1\oplus\ldots\oplus V_s$ and  $V=U_1\oplus\ldots
\oplus U_t$ be the corresponding decompositions into a sum of eigenspaces. Then  
$V^\ast=U_1^\ast\oplus\ldots\oplus U_t^\ast$ is the eigenspaces decomposition 
for $l\mapsto lb^{-1}$, and the eigenvalue corresponding to $U_j^\ast$ is $\mu_j^{-1}$. 

Next, we have 
$$ M=V\ot V^\ast=\bigoplus_{1\leq i\leq s,\ 1\leq j\leq t} V_i\ot U^\ast_j\,,$$
each of the summands $V_i\ot U^\ast_j$ being invariant under $\Phi\col x\mapsto axb^{-1}$, 
with the eigenvalue $\lambda_i\mu_j^{-1}$. Hence the set $\Sigma$ 
of eigenvalues of $\Phi$ on $M$ is the set of all distinct numbers of the form 
$\lambda_i\mu_j^{-1}$, and the eigenspace correponding to $\sigma\in\Sigma$ is 
$$ M_\sigma=\bigoplus_{(i,j)\in S_\sigma} V_i\ot U^\ast_j\,,$$
where  
$$ S_\sigma=\{(i,j)\mid 1\leq i\leq s,\ 1\leq j\leq t,\ \lambda_i\mu_j^{-1}=\sigma\}.$$

Since $M=\oplus_{\sigma\in\Sigma}M_\sigma$, it follows that  
$$ M\ot M\ot M=\oplus_{\sigma\in\Sigma}M_\sigma\ot M\ot M\,.$$
Let $\cT_\sigma$ be the component of $\cT$ in $M_\sigma\ot M\ot M$. 

For any tensor $w_i=x_i\ot y_i\ot z_i$ which is a member of $\cP$, the $x_i$ is an 
eigenvector for $\Phi$, that is $x_i\in M_\sigma$ for some $\sigma$. Therefore $\cT_\sigma$ 
is the sum of all $w_i$ such that $x_i\in M_\sigma$. There are at least $\rk(\cT_\sigma)$ of 
such $w_i$. Hence we obtain the inequality  
$$ l\geq\sum_{\sigma\in\Sigma}\rk(\cT_\sigma)\,.$$

Observe that if $(i,j),(i',j')\in S_\sigma$ and are distinct, then $i\ne i'$, $j\ne j'$. So 
we can apply Proposition~4 to compute the rank of $\cT_\sigma$ (with appropriate renumbering of 
the spaces  $V_i$ and $U_j$), and obtain 
$$ \rk(\cT_\sigma)=\sum_{(i,j)\in S_\sigma} \rk(\la d_i,f_j,3\ra)\,,$$
where $d_i=\dim V_i$ and $f_j=\dim U_j$. As any pair $(i,j)$ corresponds to some  $\sigma$, 
we see that      
$$ \sum_{\sigma\in\Sigma} \rk(\cT_\sigma)=\sum_{1\leq i\leq s,\ 1\leq j\leq t} 
  \rk(\la d_i,f_j,3\ra)\,.$$
It remains to show that the latter  sum is $\geq23$. 

Obviously, $\ov d=\{d_1,\ldots,d_s\}$ and  $\ov f=\{f_1,\ldots,f_t\}$ are partitions of the 
number $3$, and $s\geq2$, because $a$ is not a scalar matrix. There are three partitions of $3$: $3$, 
$21$, and $111$, where we denote for brevity $21=\{2,1\}$, $111=\{1,1,1\}$. If $\ov d=21$ 
and $\ov f=3$, then $l\geq\rk(\la2,3,3\ra)+\rk(\la1,3,3\ra)\geq14+9=23$. If $\ov d=21$ 
and $\ov f=21$, then $l\geq\rk(\la2,2,3\ra)+\rk(\la2,1,3\ra)+\rk(\la1,2,3\ra)+\rk(\la1,1,3\ra)
=11+6+6+3=26$. And if $\ov d=111$, then always $\rk(\la d_i,f_j,3\ra)=\rk(\la1,f_j,3\ra)=3f_j$, 
whence $\sum_{i,j}\rk(\la d_i,f_j,3\ra)=3\cdot\sum_j 3f_j=9\sum_jf_j=27$. In the case 
$\ov d=21$, $\ov f=111$ we can argue in the similar way. 

The proof of Proposition 5, and so of Theorem 2, is complete.

\paragraph{5. Finiteness of the set of candidates.} 
Let $\cP$ be a hypothetical decomposition of length $\leq22$ for $\cT$. It follows from Theorem 2
that there are only finitely many possibilities for the isomorphism class of $\Aut(\cP)$. 
However, this does not give a guarantee that there are finitely many poosibilities for $\Aut(\cP)$, 
because for a given finite subgroup $X\leq\Gamma(\cT)$ the group $\Gamma(\cT)$ can contain, in general,
infinitely many subgroups isomorphic to $X$ (say, all subgroups conjugate with $X$). 

It is easy to see, however, that if $X\leq\Gamma(\cT)$ is a subgroup, $Y=gXg^{-1}$ is a conjugate to it, 
and $\cA$ is an $X$-invariant decomposition for $\cT$, then $\cB=g\cA$ is a $Y$-invariant decomposition 
for $\cT$ (and, conversely, every $Y$-invariant decomposition of $\cT$ is $\cB=g\cA$, where 
$\cA$ is an $X$-invariant decomposition). So when studying the decompositions of $\cT$ which are 
invariant under finite subgroups we can restrict our attention and to consider a unique subgroup from 
each conjugacy class of subgroups.  

We have mentioned already that $\Gamma(\cT)\cong PSL(3,\mC)^{\times3}\sdir Q$, where $Q\cong S_3$. 
The group $PSL(3,\mC)^{\times3}=\Gamma^0(\cT)$ is an algebraic group, and 
it is easy to see that the conjugation by an element of $Q$ acts on $\Gamma^0(\cT)$ as a polynomial 
map. Therefore $\Gamma(\cT)$ is a (non-connected) algebraic group. But it is well known that if $G$ 
is an algebraic group over an algebraically closed field of characteristic $0$, and $X$ is any 
finite group, then $G$ contains only finitely many conjugacy classes of subgroups isomorphic 
to $X$. (See, e.g., \cite{Slodowy}, Theorem 1, or \cite{Platonov}, Ch.2, Theorem 17. Actually, 
in these sources stronger statements are proved.) Thus, there are only finitely many possibilities 
for $\Aut(\cP)$, up to cojugacy in $\Gamma(\cT)$.

\paragraph{6. Further restrictions.} 

It is clear that the set of conjugacy classes of subgroups of $\Gamma(\cT)\cong PSL(3,\mC)^{\times3}
\sdir S_3$ that are isomorphic to a subgroup of $S_{22}\times S_3$ is very large. 
To give an observable description of this set is a technically difficult task by itself. The author 
thinks that this set well may contain billions of groups ! So we should obtain further restrictions 
on possible $\Aut(\cP)$, say to show that this group can not contain  certain elements or subgroups. 

The aim of this section is to prove the following statement. 

\noindent{\bf Theorem 8.} {\em If $\cP$ is a decomposition of length $\leq22$ for $\cT=\la 3,3,3\ra$, 
then $\Aut(\cP)$ does not contain elements of the form $T(a,E,E)$, where $a\ne E$ 
(and the elements $T(a,b,c)$ such that exactly one of $a,b$, and $c$ is different from $E$).}
\medskip

(Note that the statement in parentheses is a trivial corollary of the first one. )

To prove Theorem 8 it is suficient to prove the following proposition. 

\noindent {\bf Proposition 9.} {\em Let $g=T(a,E,E)$, where $a\ne E$, let $w$ be an arbitrary 
decomposable tensor, and $\{w,gw,\ldots g^{l-1}w\}$ be its orbit under cyclic group $\la g\ra$. 
Then there exist decomposable tensors $w_1,\ldots,w_k$ such that $w_1+\ldots+w_k=w+gw+\ldots+g^{l-1}w$, 
$k\leq l$, and all $w_i$ are $g$-invariant. }

Indeed, suppose that $\cP$ is a $g$-invariant decomposition of $\cT$. For each  
$\la g\ra$-orbit $\cO\sse\cP$ there exists a set of decomposable tensors $\cO'$ such that  
$|\cO'|\leq|\cO|$, all elements of $\cO'$ are $g$-invariant, and the sum of elements of $\cO'$ 
is equal to the sum of elements of $\cO$. Replacing all $\cO$'s by $\cO'$, we obtain aa set of 
decomposable tensors $\cP'$ such that the sum of elements of $\cP'$ is $\cT$, all elements of 
$\cP'$ are $g$-invariant, and $|\cP'|\leq|\cP|$. But then $|\cP'|\geq23$ by Proposition~5, whence 
$|\cP|\geq23$, a contradiction. 

\noindent {\bf Lemma 10.} {\em Suppose $g=T(a,b,c)$ is of finite order $m$. Then $g$ can be 
represented as $g=T(a_1,b_1,c_1)$, where the order of each of $a_1$, $b_1$, and $c_1$ divides 
$m$ (or equals $m$). }
\medskip

{\em Proof.} We have $\id_{M\ot M\ot M}=g^m=T(a^m,b^m,c^m)$, whence $a^m=\lambda E$, $b^m=\mu E$, 
$c^m=\nu E$. Put $a_1=\lambda^{-1/m}a$, $b_1=\mu^{-1/m}b$, $c_1=\nu^{-1/m}c$. Then 
$T(a,b,c)=T(a_1,b_1,c_1)$, and moreover $a_1^m=b_1^m=c_1^m=E$. \hfill $\square$
\medskip

Now begin to prove Proposition 9. When $l=1$ the statement is trivial, so below we assume $l>1$. 
Let $m$ be the order of $g$. Then $l$ divides $m$. Moreover, we can assume, by Lemma 10, that 
$g=T(a,E,E)$ and the order of $a$ is $m$. 

Let $\lambda_1,\ldots,\lambda_t$ be all the distinct eigenvalues of $a$. Then the eigenvalues 
of the map $A\col x\mapsto ax$ of $M$ are the same $\lambda_i$, the multiplicity of $\lambda_i$ 
on $M$ equals thrice its multiplicity in the usual sense. Also, the eigenvalues of 
$B\col z\mapsto za^{-1}$ are $\lambda_i^{-1}$, and the multiplicity of $\lambda_i^{-1}$ is again 
thrice the multiplicity of $\lambda_i$. Thus, we have decompositions 
$$ M=K_1\oplus\ldots\oplus K_t\,,\qquad M=L_1\oplus\ldots\oplus L_t\,,$$
where  
$$ K_i=\{x\in M\mid ax=\lambda_ix\}\,,\qquad L_i=\{z\in M\mid za^{-1}=\lambda_i^{-1}z\}\,.$$  

Let $w=x\ot y\ot z$ be the decomposable tensor as in the hypothesis of the Proposition. 
We have $g^iw=a^ix\ot y\ot za^{-i}$. Decompose  
$$ x=x_1+\ldots+x_t\,,\quad x_i\in K_i\,,\qquad z=z_1+\ldots+z_t\,,\quad z_i\in L_i\,.$$
Hence 
$$ w=\sum_{i,j=1}^t x_i\ot y\ot z_j\,.$$ 
Notice that $g$ acts on the subspace $K_i\ot M\ot L_j$ by the multiplication by $\lambda_i
\lambda_j^{-1}$. The latter equals $1$ if $i=j$, and is a nontrivial $m$-root of $1$ if $i\ne j$. 
Hence  
$$ \sum_{k=0}^{m-1}g^kw=m\sum_{i=1}^tx_i\ot y\ot z_i\,,$$ 
whereas the summands $x_i\ot y\ot z_j$ with $i\ne j$ give zero when summed over $\la g\ra$.

Some of $x_i$ or $z_i$ may vanish. We assume, up to renumbering, that $x_i\ne0$ and $z_i\ne0$ 
when $i\leq s$, and $x_i=0$ or $z_i=0$ when $i>s$. At last, note that since $g^lw=w$, we have     
$$ \sum_{i=0}^{l-1}g^iw=(l/m) \sum_{i=0}^{m-1}g^iw\,.$$
Thus, the orbit sum for $w$ is equal to $l\sum_{i=1}^s x_i\ot y\ot z_i$. 

All the tensors $x_i\ot y\ot z_i$ are $g$-invariant. So the orbit sum is the sum of 
$s$ \ $g$-invariant decomposable tensors. This proves the proposition when $s\leq l$. 

Obviously, $s\leq t\leq3$. As $l\geq2$, the only possible case where $s>l$ is the case $s=3$, $l=2$. 
Take this case to a contradiction. We have     
$$ x=x_1+x_2+x_3\,,\qquad ax=\lambda_1x_1+\lambda_2x_2+\lambda_3x_3\,, $$
$$a^2x=\lambda_1^2x_1+\lambda_2^2x_2+\lambda_3^2x_3\,. $$

Since $\lambda_i$ are pairwise distinct, and $x_i$ are linearly independent, it follows from 
Wandermonde that $x$, $ax$, and $a^2x$ are independent. Whence $g^2w\ne w$, a contradiction. 

The proof of Proposition 9, and so of Theorem 8, is complete.

\end{document}